\documentclass{article}

\usepackage[final]{neurips_2024}

\usepackage[utf8]{inputenc} % allow utf-8 input
\usepackage[T1]{fontenc}    % use 8-bit T1 fonts
\usepackage{hyperref}       % hyperlinks
\usepackage{url}            % simple URL typesetting
\usepackage{booktabs}       % professional-quality tables
\usepackage{amsfonts}       % blackboard math symbols
\usepackage{nicefrac}       % compact symbols for 1/2, etc.
\usepackage{microtype}      % microtypography
\usepackage{xcolor}         % colors

\usepackage{booktabs}
\usepackage{graphicx}
\usepackage{caption}

\title{Generative AI Regulation Can Learn From Social Media Regulation}

\author{
  Ruth~E.~Appel \\
  Department of Communication\\
  Stanford University\\
  Stanford, CA 94305 \\
  \texttt{rappel@cs.stanford.edu} \\
}

\begin{document}

\maketitle

\begin{abstract}
  There is strong agreement that generative AI should be regulated, but strong disagreement on how to approach regulation. While some argue that AI regulation should mostly rely on extensions of existing laws, others argue that entirely new laws and regulations are needed to ensure that generative AI benefits society.
  In this paper, I argue that the debates on generative AI regulation can be informed by the debates and evidence on social media regulation. For example, AI companies have faced allegations of political bias regarding the images and text their models produce, similar to the allegations social media companies have faced regarding content ranking on their platforms.
  First, I compare and contrast the affordances of generative AI and social media to highlight their similarities and differences. Then, I discuss specific policy recommendations based on the evolution of social media and their regulation. These recommendations include investments in: efforts to counter bias and perceptions thereof (e.g., via transparency, researcher access, oversight boards, democratic input, research studies), specific areas of regulatory concern (e.g., youth wellbeing, election integrity) and trust and safety, computational social science research, and a more global perspective.
  Applying lessons learnt from social media regulation to generative AI regulation can save effort and time, and prevent avoidable mistakes.
\end{abstract}

\section{Introduction}

When Google's generative AI model Gemini produced images of racially diverse Nazis in early 2024, it led to a public outcry and allegations of anti-conservative bias \citep{Robertson2024}. Almost a decade earlier, the first allegations of anti-conservative bias were made against social media platforms like Facebook \citep{Barrett2021}, and have continued to persist e.g. during Senate hearings \citep{Romm2019} and when former President Trump was banned from Twitter (now X) and Facebook \citep{Barrett2021}. This shows that the content moderation challenges that emerging technologies face are not entirely new. Media scholars have called attention to the fact that new technologies often elicit similar questions and concerns as their predecessors \citep{Wartella1985}. Generative AI is the latest technology to garner widespread attention and raise societal and regulatory concerns, but so have social media and other technologies before it.

The aim of this paper is to show that what has been learnt with regards to social media regulation in the past two decades can inform the regulation of generative AI going forward. While there is strong agreement that generative AI should be regulated, there is strong disagreement on how to approach regulation. Some argue that AI regulation should mostly rely on extensions of existing laws \citep{Huttenlocher2023}, while others argue that entirely new laws and regulations are needed and have proposed laws and regulations such as the EU AI Act, the White House Executive Order on AI, or California's AI Safety Bill SB 1047. Analyzing the evolution of social media regulation can provide insights into which approaches to regulation are promising when it comes to generative AI, which in turn can save effort and time, and prevent avoidable mistakes.

The focus of this paper is on content moderation, i.e. how to design and regulate the content that is generated by a generative AI model or shown on a social media platform. Further, the paper focuses on regulation in a broad sense, which can include self-regulation of industry players to ensure harmless output or avoiding bias, and formal laws such as the General Data Protection Regulation or the White House Executive Order on AI. The first section compares and contrasts the affordances of generative AI and social media to highlight their similarities and differences. The second section discusses specific policy recommendations based on the evolution of social media and their regulation.

\section{Affordances of generative AI and social media}

To shed light on the similarities and differences between specific media, we can analyze their affordances. Affordances are the features that characterize a medium. Both generative AI, e.g. in the form of a chatbot like OpenAI's ChatGPT or Anthropic's Claude, and social media, e.g. in the form of Meta's Facebook or X (formerly Twitter), can be considered media that allow to create and distribute content and are shaped by specific features. The features discussed here pertain to a medium in general, but may not apply to every instance, that is, a specific generative AI application or social media platform may differ from the norm in terms of its affordances.

Based on an analysis of commonly used generative AI applications (e.g., ChatGPT and Claude) as well as social media applications (e.g., Facebook and X), I identified key features that generative AI and social media share or that differentiate them. The analysis of features is grounded in work by \cite{Clark1996}, who discusses several features of media that fall into three categories: medium, control, and immediacy. Since  \citeauthor{Clark1996}'s features were originally meant to capture affordances of face-to-face communication,\footnote{There are contextual differences between face-to-face communication on the one hand and generative AI and social media on the other, such as where and why they may be used. This paper focuses on the comparison of generative AI and social media, and therefore focuses on features in \citeauthor{Clark1996}'s model that are pertinent to generative AI and social media, but not the comparison to other media.} I added new features and removed features that are less relevant to the comparison of generative AI and social media. I will point out each feature that is adapted from Clark.

I will first address why social media are an imperfect analogy to generative AI, and then outline why this analogy can still be very helpful to highlight key features that can inform regulation.

\begin{table}[h]
\centering
\caption{Comparison of affordances of generative AI and social media}
\label{tab:afford}
\resizebox{\textwidth}{!}{%
\begin{tabular}{llcc}
\toprule
Feature & Definition & Generative AI & Social Media \\
\midrule
\multicolumn{4}{l}{\textit{Medium}} \\
\quad Spatial separation & Content is generated in different locations & \textbf{Yes} & \textbf{Yes} \\
\quad Direct connection & Medium is conversation partner & Yes & No \\
\quad User connections & Medium connects user to other users & No & Yes \\
\quad Dialogue-by-default & Actions occur in a dialogue & Yes & No \\
\quad Recording & User actions are recorded & \textbf{Yes} & \textbf{Yes} \\
\quad Personalization & User context and preferences are learnt over time & \textbf{Yes} & \textbf{Yes} \\
\quad Single output & Medium presents usually just a single output & Yes & No \\
\quad Infinite content & Content is served infinitely & No & Yes \\
\quad General content & Content can pertain to any domain & \textbf{Yes} & \textbf{Yes} \\
\quad General purpose & Medium serves many functions & Yes & No \\
\quad Use of AI & Medium learns patterns from data & \textbf{Yes} & \textbf{Yes} \\
\quad Abstraction & Medium hides its complexity & \textbf{Yes} & \textbf{Yes} \\
\quad Black-box & How algorithmic decisions are made is intransparent & \textbf{Yes} & \textbf{Yes} \\
\multicolumn{4}{l}{\textit{Control}} \\
\quad Content moderation & Content is moderated at all & \textbf{Yes} & \textbf{Yes} \\
\quad Invisible content moderation & Most content moderation is not visible to the user & Yes & No \\
\quad Content moderation pre-generation & Content is moderated before it is received by the user & Yes & No \\
\quad Self-determination & User can decide themselves how to act & \textbf{Yes} & \textbf{Yes} \\
\quad Self-expression & User can express themselves & \textbf{Yes} & \textbf{Yes} \\
\quad Simultaneity & User can receive and produce content concurrently & No & Yes \\
\multicolumn{4}{l}{\textit{Immediacy}} \\
\quad Instantaneity & Actions are perceived almost immediately & \textbf{Yes} & \textbf{Yes} \\
\quad Evanescence & Medium quickly recedes to the background & \textbf{Yes} & \textbf{Yes} \\
\bottomrule
\end{tabular}}%
\vspace{-5pt}
\caption*{\footnotesize{\textit{Note}: The features spatial separation, recording, self-determination, self-expression, simultaneity, instantaneity, and evanescence, as well as the categories medium, control and immediacy are based on \cite{Clark1996}. Instances where features of generative AI are similar to features of social media are highlighted in bold.}}
\end{table}

\subsection{Generative AI and social media are not perfectly comparable}

By definition, an analogy is not a perfect match, otherwise the objects of comparison would be the same thing. As Jacob Stern puts it: ``[T]his is just the nature of analogies: They are illuminating but incomplete'' \citep{Stern2023}.

Table~\ref{tab:afford} reveals differences in affordances between generative AI and social media. With regards to features of the medium, generative AI and social media show some variation. While generative AI such as ChatGPT constitutes a conversation partner for the user and interacts in a dialogue with the user, social media are merely mediating between the user and their human conversation partner (e.g., when a social media algorithm displays one user's post on another user's feed) and tend to involve a sequence of one-off actions. Relatedly, while social media foster direct connections between users, generative AI is usually used by a single person at a time.
While generative AI tends to respond to prompts, usually with a single output instead of multiple outputs, and does not continue to serve content unless the user requests it, social media often feature infinite scroll or auto-play that serve content as long as the user is on the platform. The purpose of social media tends to be limited to social communication, while generative AI is considered a general purpose technology that could serve various functions, including as a text writer or reviewer, a calculator, a programmer and much more. 

With regards to control features, a feature \cite{Clark1996} proposed is simultaneity, which is the user's ability to receive and produce content concurrently. Simultaneity is given for social media --- e.g., one user might send a message at the same time as another user is sending them a message ---, but not for generative AI, which operates in a sequential dialogue of user input and model output. Important differences between generative AI and social media concern content moderation: Even though both generative AI and social media feature content moderation, content moderation in generative AI tends to be less visible than on social media. Social media platforms may occasionally take hardly visible actions such as downranking posts, but many social media content moderation actions such as removal of a post or user are clearly visible. Generative AI models, on the other hand, are built and fine-tuned to moderate content in a certain way (e.g., to avoid providing dangerous information) without the user necessarily becoming aware of the moderation. Generative AI content moderation may be invisible to the user because the model will usually respond, and not necessarily provide a reason if it refuses to respond to a prompt directly, which makes moderation less obvious than a missing response or a refused response citing the reason for refusal. Relatedly, generative AI models tend to moderate \textit{before} the content is shown to the user, e.g. by refusing to reply to a prompt, while social media content moderation tends to occur only \textit{after} content made it onto a platform, e.g. when a post was reported as harmful misinformation.

Beyond specific features of generative AI and social media, there are differences in their context and potential consequences. In terms of business model, most social media companies rely on revenue from advertisements, while prominent generative AI companies have so far leaned towards freemium subscription models. 
While the potential harm of social media to democracy and society has been an important focus of scholarly and public attention \citep{Persily2020}, some argue that the destructive potential of AI may be at another level since it may present a larger threat or stronger geopolitical advantage \citep{Stern2023}. 
Generative AI and social media differ also in the level of uncertainty they bring. For example, auditing and discovering vulnerabilities in systems that are probabilistic \citep{Cattell2024}, like generative AI models, implies new complexities that traditional, deterministic social media algorithms do not entail.
Finally, generative AI and social media may differ in areas that have so far remained legally uncertain, such as questions of liability (e.g., for harms results from media use) and copyright.

\subsection{Generative AI and social media are comparable in key aspects}

The analogy between generative AI and social media is valuable despite its imperfection because their key features are similar. Importantly, the shared affordances of generative AI and social media imply that both of these media necessarily moderate content and thus face complex content moderation challenges and public scrutiny.

Table~\ref{tab:afford} shows key similarities between generative AI and social media when it comes to the features of each medium. Both generative AI and social media allow for spatial separation, that is, the conversation partners usually generate content in different physical spaces --- e.g., in a home office and at a data center for generative AI --- and are not copresent (copresence is one of the features of face-to-face communication in \cite{Clark1996}). Both generative AI and social media are recording user data (the recording feature is adapted from \citeauthor{Clark1996}'s recordlessness feature). Both media can learn about a user's context and their preferences over time to personalize their output, e.g. by updating the chatbot's memory or personalizing a recommendation algorithm. Further, both generative AI and social media can feature content on all kinds of domains (e.g., hobbies, jobs, politics). Both are powered by artificial intelligence (AI), that is, they rely on learning patterns from data to perform well on tasks such as generating text or recommending content, although generative AI relies on more recent deep learning models while social media tends to rely on traditional machine learning approaches such as recommender systems. Both media also feature abstraction, that is, they hide the complex technical implementation details from the user behind a simple user interface. Further, generative AI and social media algorithms tend to be black-box, that is, algorithmic decisions are intransparent --- almost always for users, but often also for experts because mechanistic interpretability that can explain why a deep learning model made a certain decision is in its infancy.

With regards to control features \citep{Clark1996}, both generative AI and social media feature content moderation, that is, the medium shapes what content is allowed to appear. Both media also meet \citeauthor{Clark1996}'s criteria for self-determination, i.e. a user's ability to decide themselves how to act, and self-expression, i.e. a user's ability to express themselves on a medium.

With regards to immediacy \citep{Clark1996}, both generative AI and social media share instantaneity  \citep{Clark1996}, i.e. that actions are perceived almost immediately, and evanescence  \citep{Clark1996}, i.e. that the medium recedes to the background quickly once it is not actively used anymore.

Beyond features, the evolution of generative AI is similar to the evolution of social media in that both are characterized by limited, lagging regulation and large inflows of money \citep{Stern2023}.

\section{Learnings from social media regulation for generative AI regulation}

As the review of the affordances has shown, generative AI and social media share important features, including the use of AI and content moderation. Although generative AI and social media differ on some dimensions, these differences suggest, for the most part, differences in degree, and not differences in kind when it comes to regulation. Thus, lessons learnt from social media regulation may be relevant to generative AI regulation. This paper addresses four policy recommendations for generative AI regulation based on the evolution of social media regulation: (1) counter bias and perceptions thereof (e.g., via transparency, oversight boards, researcher access, democratic input, multidisciplinary research), (2) address specific regulatory concerns (e.g., youth wellbeing, election integrity) and invest in trust and safety, (3) promote computational social science research, and (4) take on a more global perspective.

\subsection{Counter bias and perceptions thereof}

Given that both generative AI and social media share the key features content moderation, use of AI, that they are black-box and abstract the complexity of algorithmic decision-making away such that much of the decision-making is intransparent, it is no surprise that both generative AI companies and social media companies have faced allegations of bias, including allegations of anti-conservative political bias \citep{Robertson2024,Barrett2021}. While there is no evidence of anti-conservative bias for social media \citep{Barrett2021}, multiple studies have shown political bias in generative AI. For example, compared to representative opinion polls, large language models were found to output biased opinions \citep{Durmus2023,Santurkar2023}, and multiple studies showed left-leaning bias in generative AI such as ChatGPT \citep{Rozado2023,Rottger2024}.

Generative AI models have also been shown to exhibit other forms of bias, such as anti-Muslim bias \citep{Abid2021}, bias towards Western culture \citep{Naous2023}, and stereotypical depictions of race, gender, age, nationality, and socioeconomic status \citep{Nangia2020}.

Addressing such biases is as important as it is challenging. It is important to address biases because biases can harm users by leading to lower-quality output, they can entrench historical biases and stereotypes, and they can undermine trust in model developers, model deployers, and regulators of generative AI. It is challenging to address biases because they are challenging to measure accurately (e.g., they may be sensitive to the specific prompt design \citep{Rottger2024}) and because it is not clear where exactly biases stem from. Biases can arise at different points in the development and deployment of generative AI, including training and data curation, fine-tuning, evaluation and feedback, real-time moderation, customization and control of models \citep{Suresh2019,Ferrara2023}. 

Social media companies have taken different approaches to address biases or perceptions thereof that mainly focus on transparency about algorithms and decision-making, gathering input from users and learning from case studies, and increasing user choice.

\subsubsection{Increase transparency and researcher access}

The features content moderation, use of AI, black-box and abstraction also give rise to transparency challenges for social media and generative AI.
Generative AI transparency has been poor as shown in the Foundation Model Transparency Index \citep{Bommasani2023,Bommasani2024}. Social media companies have pursued multiple different approaches to increase transparency and generative AI can learn from this playbook. For example, Facebook's parent company Meta introduced features such as ``Why am I seeing this ad?'' that allowed users to understand why they were served certain ad content \citep{Thulasi2019}, created blog posts and a Transparency Center providing some information on the role of AI and other factors in content recommendation \citep{Clegg2023,Metaa}, and established an independent oversight board of experts that adjudicates particularly contentious content moderation decisions \citep{meta2024oversightboard}. These initiatives do not come without problems. In response to the launch of Facebook's oversight board, ``The real Facebook Oversight Board'' was created, which brought experts together to argue for more independence, transparency and regulation \citep{TheRealFacebookOversightBoard2022}.

An important aspect of transparency is allowing for third-party evaluations. Efforts to create APIs accessible to researchers, such as the Facebook Open Research and Transparency Researcher API and the TikTok Research API, or to design academic-industry collaboration such as the Facebook and Instagram Election Study are helpful but imperfect \citep{Wagner2023}. The Coalition for Independent Technology Research was founded after researchers at different institutions faced difficulty maintaining or gaining access to social media data for research purposes \citep{CoalitionforIndependentTechnologyResearch2022}. Importantly, we can learn from these shortcomings. Researcher access programs to evaluate technology should be characterized by sufficient resources (including staffing, infrastructure, and funding), incentives that are compatible with academic research (e.g., data retention policies, persistent API access and publication permission for researchers), sound knowledge sharing processes between internal and external researchers to help understand data availability and feasibility, helpful documentation, privacy preserving measures (e.g., aggregation of user data) and timeliness in terms of publication or addressing of issues that researchers discovered. To protect researchers involved, researchers have called for ``safe harbors,'' that is, legal protection for researchers pursuing legitimate research purposes, initially for social media \citep{Abdo2022} and more recently for generative AI \citep{pmlr-v235-longpre24a}.
Additional proposals to facilitate external generative AI research include data donations \citep{Sanderson2024}. 

Regulations like the Digital Services Act prescribe transparency by requiring audits of social media companies \citep{EuropeanCommission2023}, and similar auditing efforts are imaginable for generative AI. In fact, some scholars suggest to extend and adapt DSA rules for social media platforms to generative AI \citep{Hacker2023}.

While the specific implementation of these transparency efforts may be contentious and requires nuance, ideas such as short and accessible explanations of the technology, independent oversight mechanisms, researcher access and mandatory audits are viable options for increasing transparency via generative AI regulation.

\subsubsection{Gather democratic input to inform technology}

Generative AI and social media share features that make them complex, including that the content they feature can pertain to a variety of domains, that there is potential for personalization, and that content could be moderated in various different ways. Given the vast set of choices that developers face, one approach is to gather input directly from users to determine what a good system may look like.
In terms of gathering input from users to enable democratic decisions about the nature of regulation and content moderation, different initiatives have been launched over the past few years to deliberate issues ranging from cyberbulling on social platforms to the rules and constitutions that inform generative AI models \citep{Wetherall-Grujic2023}, relying on the much older idea of deliberative democracy \citep{Eagan2016}. Social media also offers case studies of networks where content moderation seems to be broadly accepted and deliver productive results, such as in the case of the deliberation platform vTaiwan \citep{Miller2019} or a neighborhood-focused social network \citep{Oremus2024}. Finally, social media researchers have studied how to embed important societal values into AI \citep{Bernstein2023}, which could also inform how such values can be embedded into generative AI.

\subsubsection{Promote user choice}

Another option to empower users to make choices in the face of features like content moderation and the varied nature of content is to enable users to set up rules for a subset of the system.
The social media platform Mastodon is a prominent example in terms of increasing user choice in such a way. Mastodon is built on the idea that different communities can create their own servers and set and enforce their own content moderation rules \citep{Mastodon2024}. This highlights that the feature of personalization may be a potential route for resolving content moderation dilemmas. Content moderation questions with regards to generative AI and social media are similar and it is not clear what opinion representation should be the default \citep{Redpoint}. This suggests that increased personalization of models may be an answer \citep{Redpoint}.

\subsection{Address specific regulatory concerns and invest in trust and safety}

The feature of content moderation that generative AI and social media share comes with especially thorny issues such as preventing the spread of harmful misinformation and protecting user wellbeing.
Social media companies have invested in teams that address these specific regulatory concerns. Examples include teams at companies like Google, Meta and Microsoft working on youth wellbeing and mental health in general, election integrity, preventing spam, preventing the spread of child sexual abuse material, preventing harmful misinformation, detecting deceptive campaigns, and ensuring trust in the platform and safety of its users overall.

Generative AI chatbots have already been rated on AI-related principles that apply just as much to social media. Common Sense Media published rankings of different generative AI models with regards to the following principles: put people first, prioritize fairness, be trustworthy, keep kids and teens safe, be effective, help people connect, use data responsibly, and be transparent and accountable \citep{CommonSenseMedia2024}. Yet, generative AI companies do not have teams at the same scale as social media companies to address these issues.

Generative AI companies are much smaller and younger than some of the social media giants, thus it is not surprising that they do not have as much dedicated staff to work on these issues. Going forward, however, adding diverse staff beyond engineers that can bring in expertise to address issues such as user mental health or combating misinformation seems important. Investment in trust and safety teams seems particularly crucial, and it is encouraging to see that companies like OpenAI and Anthropic are investing in this area, with OpenAI publishing the first-ever report on the activity of deceptive campaigns on generative AI platforms in May 2024 \citep{Nimmo2024}. The policies social media companies have put in place to decide how and when to moderate individual users, and the best practices they have developed to uncover abuse such as deceptive campaigns that try to interfere with elections or spam users, could inform the approaches generative AI companies take. This includes developing a repertoire of content moderation approaches, which could include bans, but also more cautious interventions such as warnings and strikes for misbehavior, putting more guardrails in place or throttling usage for users that tried to abuse generative AI models in the past. Social media companies also gained experience in involving the user community in content moderation decisions (e.g., in the case of BirdWatch \citep{Wojcik2022}) and how to collaborate across platforms, and generative AI companies could consider how these approaches could be adapted to their platforms.

\subsection{Promote computational social science research}

Both generative AI and social media allow users to express themselves and allow for a connection, be it to other users or to an AI with a vast pool of knowledge. These features suggest that both of these media are so important and powerful because of how they interact with users. They are neither purely technical, nor purely social systems. This suggests that multidisciplinary study --- computational social science --- is needed to understand, evaluate and shape these systems.

In fact, the recommendations above, whether regarding measures to reduce bias or enhance user wellbeing, all require computational social science research to test their effectiveness. Social media companies have hired researchers from many disciplines, including computer science, psychology, political science, communication, law and others, to better understand how their platforms impact society, and how certain interventions influence society and their revenue.

Whether research is conducted in-house or via access to the platforms for external researchers, rigorous evaluations are key to ensure that media like generative AI and social media meet their goal of being helpful and not harmful to society. Further investment in research is needed because generative AI does have features that differ from previous technologies, so its impact and user preferences (e.g., with regards to privacy, personalization or content moderation) are not clear. Even the impact of previous technologies like social media has not yet been comprehensively evaluated and needs further investment. Rigorous research can inform platform and public policy when it comes to regulation, and it can enhance user trust.

This implies investing in diverse research teams that understand the interaction of humans an a given medium and that can evaluate the societal implications of a product. While AI company recruiting often focuses heavily on engineers, and some companies are more concerned with extreme risks in the more distant future, social media companies have shown the value of addressing current risks such as biases and of creating multidisciplinary teams to do so. This allows companies to test different product features and interventions effectively, e.g. to reduce spam or misinformation spread. Computational social scientists from any background, data scientists and user experience researchers would be especially helpful to address questions at the intersection of technology and humans, such as which emotional bonds may be formed between humans and AI, and what type of personalization could be implemented.

While content moderation on social media is far from a resolved issue, there is a large and growing body of academic literature that speaks to promising approaches and could inform content moderation for generative AI \citep[e.g.][]{Persily2020,Kozyreva2024}.

\subsection{Take on a more global perspective}

As the features spatial separation, general content, and use of AI imply, both generative AI and social media can be used in a variety of contexts.
Generative AI companies have grown rapidly and are serving users around the world, similar to social media companies. However, compared to social media companies, generative AI companies, at least the startups among them, seem more heavily focused on the US due to their location (with exceptions like Google DeepMind in the UK and Mistral AI in France). To address problems like biases, it is crucial that even small companies take on a global perspective and become global companies with local expertise in multiple countries. This could take the form of local offices and a focus on hiring internationally. The stakes are high. If companies fail to invest in taking user preferences and risk factors outside of the US seriously, the technology may serve large numbers of users worse (e.g., due to under-investment in non-English language content generation) and could even result in catastrophes such as promoting violence in conflict regions \citep{AmnestyInternational2022}.
Given the increasing amount of national and local regulations on generative AI, global expertise is also important to keep up with local laws.

For effective regulation, local expertise needs to be integrated into a global perspective. For example, the former Prime Minister of New Zealand suggested that a model for governing AI could follow the Christchurch Call, which is a multinational, multi-stakeholder effort bringing together governments, tech companies and civil society to eliminate violent extremist and terrorist content online \citep{Ardern2023}.

\section{Conclusion}

There are strong disagreements about the approach that should be taken to regulate generative AI. This paper argued that the regulation of generative AI can be informed by the evolution of the regulation of social media.
While social media is by far not the only analogy proposed for generative AI \citep{Maas2023}, and by no means a perfect analogy, generative AI and social media share key features that make a comparison of the two worthwhile.
Taking a close look at social media regulation efforts --- including self-regulation and laws --- reveals interesting approaches and best practices.
This paper outlined recommendations regarding transparency, researcher access, gathering democratic input, promoting user choice, addressing specific regulatory concerns, increasing investments into computational social science research, and taking on a more global perspective.
In the case of social media, self-regulation did not always work, which has resulted in multiple new laws being proposed in the past few years. These laws, but also the forms of self-regulation that were put in place, including specific approaches to increasing transparency, enhancing user choice, and investing in research, can be valuable pointers for those looking to regulate generative AI. Analyzing social media regulation may speed up the process of developing generative AI regulation. The EU AI Act may not have been able to address general purpose models as fast as it did had it not already been concerned with other forms of machine learning much earlier. Regulation takes time and effort, so where possible, resources should be saved and mistakes avoided by looking at social media regulation and research.

\begin{ack}
The author has worked on two research projects in collaboration with Meta. The author attended a Meta event where food was paid for by the company. The author is grateful to Jennifer Pan for feedback on an earlier draft.
\end{ack}

{\small
\bibliography{references}
}

\end{document}